\documentclass[showpacs,amsmath,amssymb,twocolumn,superscriptaddress,notitlepage,preprintnumbers,pra]{revtex4-1}

\usepackage[dvips]{graphicx}
\usepackage{amsmath,amssymb,amsthm,mathrsfs,amsfonts,dsfont}
 
\usepackage{dcolumn}
\usepackage{bm}
\usepackage{amsmath}
\usepackage{amssymb}
\usepackage{braket}
\usepackage{physics}
\usepackage{amsthm}
\usepackage{natbib}
\usepackage{mathtools}
\usepackage{diagbox}
\usepackage[utf8]{inputenc}
\usepackage[english]{babel}
\usepackage{scalerel}[2014/03/10]
\usepackage{stackengine}

\usepackage{hyperref}
\hypersetup{colorlinks=true, linkcolor=blue, citecolor=blue, urlcolor=black }

\DeclareMathOperator*{\E}{\mathbb{E}}
\newcommand{\cQITE}{\mathcal{\mathcal{E}}}

\newcommand{\mol}[1]{\mathrm{#1}}

\newtheorem{prop}{Proposition}

\theoremstyle{definition}

\newcommand{\bytedance}{ByteDance Ltd., Zhonghang Plaza, No. 43, North 3rd Ring West Road, Haidian District, Beijing, China}
\newcommand{\tsinghua}{Yau Mathematical Sciences Center and Department of Mathematics, Tsinghua University,  Beijing 100084, China}
\newcommand{\oxford}{Clarendon Laboratory, University of Oxford, Parks Road, Oxford OX1 3PU, United Kingdom}

\begin{document}

\title{Efficient quantum imaginary time evolution by drifting real time evolution: an approach with low gate and measurement complexity}

\author{Yifei Huang} 
\thanks{huangyifei.426@bytedance.com}
\affiliation{\bytedance}
\author{Yuguo Shao} 
\affiliation{\bytedance}
\affiliation{\tsinghua}
\author{Weiluo Ren}
\affiliation{\bytedance}
\author{Jinzhao Sun}
\thanks{jinzhao.sun@physics.ox.ac.uk}
\affiliation{\oxford}
\author{Dingshun Lv}
\thanks{lvdingshun@bytedance.com} 
\affiliation{\bytedance}

\begin{abstract}

Quantum imaginary time evolution (QITE) is one of the promising candidates for finding eigenvalues and eigenstates of a Hamiltonian. However, the original QITE proposal [Nat. Phys. 16, 205-210 (2020)], which approximates the imaginary time evolution by real time evolution, suffers from large circuit depth and measurements due to the size of the Pauli operator pool and Trotterization. To alleviate the requirement for deep circuits, we propose a time-dependent drifting scheme inspired by the qDRIFT algorithm [Phys. Rev. Lett 123, 070503 (2019)], which randomly draws a Pauli term out of the approximated unitary operation generators of QITE according to the strength and rescales that term by the total strength of the Pauli terms. We show that this drifting scheme removes the depth dependency on size of the operator pool and converges inverse linearly to the number of steps. We further propose a deterministic algorithm that selects the dominant Pauli term to reduce the fluctuation for the ground state preparation. Meanwhile, we introduce an efficient measurement reduction scheme across Trotter steps, which removes its cost dependence on the number of iterations, and a measurement distribution protocol for different observables within each time step. We also analyze the main source of error for our scheme both theoretically and numerically. We numerically test the validity of depth reduction, convergence performance, and faithfulness of measurement reduction approximation of our algorithms on $\mol{LiH}$, $\mol{BeH_2}$ and $\mol{N_2}$ molecules. In particular, the results on $\mol{LiH}$ molecule give circuit depths comparable to that of the advanced adaptive variational quantum eigensolver~(VQE) methods while requiring much fewer measurements.

\end{abstract}

\maketitle

\section{Introduction}

One of the potential killer applications of quantum computers is to compute eigenstates and eigenvalues of a problem Hamiltonian~\cite{feynman2018simulating, lloyd1996universal, aspuru2005simulated}, especially its ground state and the corresponding energy. The quantum phase estimation algorithm~\cite{kitaev1995quantum, abrams1999quantum, aspuru2005simulated} is designed to deliver the eigenvalues of a system, but it requires deep quantum circuits and many ancillary qubits, and 
remains challenging for noisy intermediate-scale quantum~(NISQ) devices~\cite{bharti2022noisy,mcardle2020quantum}.
While hybrid quantum-classical algorithms, such as variational quantum eigensolvers (VQE) and quantum approximate optimization algorithms~\cite{peruzzo2014variational,farhi2014quantum}, are flexible in terms of circuit depth and robustness to coherent error, high dimensional classical optimization and large amount of local minima may affect the performance of the algorithms for quantum many-body problems~\cite{bittel2021training,mcclean2018barren,wang2021noise}.

Quantum imaginary time evolution~(QITE)~\cite{motta2020determining, mcardle2019variational,sun2021quantum,kamakari2022digital,liu2021probabilistic,silva2021imaginary,huo2021shallow,mao2022measurement} is an alternative approach for obtaining the ground state of a Hamiltonian by mimicking the behavior of the imaginary time evolution using unitary operation on a quantum computer. The efficiency of the QITE highly relies on the
premises that (1) the Hamiltonian is local, (2) the initial state is a product state, (3) a small number of terms in the Hamiltonian. If the system is nonlocal or the initial state is an entangled state, one has to use a large Pauli basis set to ensure that the unitary approximation is valid. For chemical systems, the Hamiltonian is highly nonlocal and a full set of Pauli basis needs to be used, which scales exponentially to the system size $N$. However, one may only include the Pauli operators in the set of unitary coupled cluster with single excitations and double excitations~(UCCSD)~\cite{bravyi2002fermionic,seeley2012bravyi,romero2018strategies} into the basis set, as utilized in step-merged QITE~\cite{gomes2020efficient}. We will adopt this basis set constraint in this work as well. Even under these assumptions, the circuit depth of QITE still scales with the product of the number of Trotter steps and the number of Pauli terms in the Hamiltonian. This renders accurate quantum chemistry simulation on NISQ devices out of reach with QITE.
  
In this work, we propose an efficient algorithm that addresses the challenges in quantum imaginary time evolution. To deal with the depth requirement of QITE, we exploit the idea of random compiling named qDRIFT into QITE~\cite{campbell2019random} in Section \ref{algorithm}. A straightforward way is to sample the Trotterized imaginary time operator in the ITE, which however introduces a large simulation error. Instead, we sample by the strengths of the real time evolution operators obtained by solving the set of linear equations at each time step. In contrast the original qDRIFT proposal, which samples from a fixed distribution at each time step, we deal with a time-dependent real time evolution operator, and hence we term our main algorithm as time-dependent~(td)-DRIFT-QITE. With our scheme, we eliminate the depth dependency of QITE on the size of the Pauli set, and thus greatly reduce the circuit depth requirement. 

In Section III we introduce an efficient scheme to reduce the measurement complexity during the time evolution. As the state at each time step is only shifted slightly from the previous step by adding a layer of the circuit generated by a single Pauli operator, we can reuse the estimations of expected values from previous measurements to assist the evaluation of the new state. We show that the total measurements for an arbitrary observable are independent of the number of time steps. Moreover, we show a measurement distribution strategy for the estimation of different observables at each time step and present the number of measurements accordingly. In our algorithm, we truncate small singular values in order to maintain numerical stability, and thus leads to deviation. We show that this deviation mainly stems from the nonlocal interaction in the Hamiltonian, which has a large correlation length and large condition number. In Section IV, we provide numerical test for our algorithms on benchmark molecules including $\mol{LiH}$, $\mol{BeH_2}$ and $\mol{N_2}$.

\section{Time-dependent drifting algorithm} \label{algorithm}

Assuming a problem Hamiltonian $H$, the imaginary time evolution leads the initial state $\ket{\Psi}$ to $e^{-H\Delta t}\ket{\Psi}$ after a time slice $\Delta t$. 
To realize the imaginary time evolution on a quantum computer, Motta \textit{et al} proposed to approximate the state under the imaginary time evolution by applying a unitary operation $e^{iA\Delta t}$, which realizes the nonunitary operation $e^{-H\Delta t}$ effectively~\cite{motta2020determining}. The unitary operation can be identified by minimizing the approximation error as $\min \| \frac{1}{\sqrt{c}}e^{-H \Delta t} \ket{\Psi} - e^{iA\Delta t}\ket{\Psi} \|_2$, where $c=\bra{\Psi}e^{-2H\Delta t}\ket{\Psi}$ is the normalization factor and $\|\cdot \|_2$ is the vector norm.

Suppose that the Hermitian operator $A$ can be expanded in  the Pauli basis as $A = \sum_{i_1,...,i_D} a_{i_1,\cdots,i_D} P_{i_1, \cdots, i_D} $ with the domain size $D$.
Solving the minimization problem in terms of the coefficients $\bold{a}$~(the coefficients in the Pauli decomposition of $A$ as a vector) up to the first order of $\Delta t$, one obtains the following set of equations
\begin{equation}
\bold{S} \bold{a} = \bold{b},  \label{equation}
\end{equation}
where $\bold{S}_{ij}=\Re(\bra{\Psi}P_iP_j\ket{\Psi})$ and $\bold{b}_{j} = -{c}^{-1/2}\Im(\bra{\Psi}HP_j\ket{\Psi})$.
Here $P_i$ is the Pauli basis used to decompose $A$, and from now on the subscript denotes the label of the Pauli operator in the pool instead of the domain number. Note that we approximate the nonunitary operator at the first order, while higher order expansion can be used (see~\cite{sun2021quantum}).
We can find that the measurement cost scales exponentially to the domain size $D$ as $\exp(\mathcal{O}(D))$. To alleviate this problem, the original proposal~\cite{motta2020determining} assumed that the Hamiltonian is local, $H = \sum_{l=1}^L h_l H_l$, where $H_l$s are normalized~(Pauli operators) with signs included and $h_l$s are  positive numbers,
and considered the first-order Trotter formulation of
\begin{equation}
e^{-H\Delta t}\ket{\Psi}\approx \prod_{l} e^{-h_l H_l\Delta t}\ket{\Psi}    .
\label{trotterized_hamil}
\end{equation}
They considered to approximate the Trotterized imaginary time evolution $ e^{-h_l H_l\Delta t}$ instead of the full time-sliced  $e^{-H \Delta t}$ to bound the domain size of the unitary operators. 
In particular, if the Hamiltonian is $k$-local, the measurement cost depends polynomially on the number of Trotter steps and the number of local terms in the Hamiltonian while exponentially on the correlation length of the quantum state (see \cite{motta2020determining} for details).

The limitation of Trotterization based method is that the circuit depth scales with the number of terms in the Hamiltonian, which is typically polynomial to the system size, and hence becomes prohibitively large in the practical implementation.
To relax the depth dependence on the number of Hamiltonian, one may exploit the spirit of qDRIFT~\cite{campbell2019random} and randomized Trotterization~\cite{childs2019faster,chen2021concentration}. The key idea of qDRIFT is that one randomly samples a single term from the Hamiltonian with probability proportional to their strengths at every step and rescales the step size to match the norm of the whole Hamiltonian. Therefore, the depth of Hamiltonian simulation using qDRIFT is independent of the number of Pauli terms in the Hamiltonian.

Now we employ the idea of qDRIFT on the ansatz operators $A$ to reduce the circuit depth of QITE, which we refer to as time-dependent~(td)-DRIFT-QITE. The convergence of this drifting to the original QITE is not trivial to show given the time dependent nature of the $A$ operators at different time steps, which we will discuss later in Proposition~1. Let $A^{(j)}$ denote the $A$ operator obtained from the original QITE method at the $j$th Trotter step and the unitary as $U_j:=e^{iA^{(j)}\Delta t}$. 
Given a decomposition $A^{(j)} =\sum_{i} a_iP_i$, the quantum channel representing this randomized sampling at the $j$th step is 
\begin{equation}
\cQITE(\rho) = \sum_{i}\frac{a_i}{\norm{\bold{a}}_1}e^{iP_i\norm{\bold{a}}_1\Delta t}\rho e^{-iP_i\norm{\bold{a}}_1\Delta t}, \label{QDRIFT-QITE}
\end{equation}
where we abbreviate the superscript of $j$ for simplicity.
Similar to qDRIFT, one can show that the expression in Eq.~(\ref{QDRIFT-QITE}) agrees with the ideal unitary evolution channel $U_j\rho U_j^{\dagger}$ to the first order in $\Delta t$, and the error is of the order $\mathcal{O}(T^2 \Delta t)$, which is the same as the first-order Trotter method. The difference is that the generators of the real time evolution at every step in QITE are different, while they are the same for Hamiltonian simulation of dynamics. This time-dependence of $A$ has a similar spirit to continuous qDRIFT~\cite{berry2020time}, while the difference is that our algorithm is implemented with discrete time dependence.

The following two properties of td-DRIFT-QITE make it ideal for imaginary evolution of molecular systems. First, due to long-range interactions between orbitals, molecular systems generally requires a large operator pool, for instance, the UCCSD operator pool with the size of the order $\mathcal{O}(N^4)$. 
Instead, our scheme is independent of the size of the operator pool when we sample the $A$ operators. As the expectation value of the Hamiltonian approaches the ground state energy exponentially fast along the ITE path, the total evolution time required for ITE may not be large. It is worth to remark that the idea of randomization in QITE was initially mentioned and tested in Ref.~\cite{cao2021quantum}, whereas their scheme randomly shuffles the order of Hamiltonian terms in each Trotter steps and hence the circuit depth still depends polynomially on $L$.

More importantly, we show that the convergence rate is proportional to the inverse of the time step number under mild condition of the convergence property of the original QITE.
We define the ideal unitary trajectory that approximates the imaginary time evolution as $\Xi_{j}=U_j...U_1$, and we define a random variable $X^{(j)} = \norm{\bold{a}}_1P_i$ that is drawn from the decomposition of $A^{(j)}$ with probability $p_{j,i} = \frac{a_i}{\norm{\bold{a}}_1}$ and the drifted unitary at the $j$th step $V_j := e^{iX^{(j)}\Delta t}$. 
The randomized drifted unitary operation at the $j$th step is represented as $\Lambda_j = \E(V_j|\Lambda_{j-1})\Lambda_{j-1}$  where  the drifted unitary is conditioned on the previous unitary $\Lambda_{j-1}$ and the initial condition $\Lambda_1=\E(V_1)$.
We have the following results to ensure the convergence of the algorithm.
\begin{prop}
The approximation error of td-DRIFT-QITE is bounded by the inverse of iteration $K$ as
$
\norm{\Lambda_K-\Xi_K}\leq \mathcal{O}(K^{-1})  
$
where $\norm{.}$ is the spectral norm and the total evolution time is fixed. \label{prop:bound_prop}
\end{prop}
This result indicates that the convergence rate is proportional to the inverse of the iteration number, and we refer to Appendix~\ref{bound_proof} for detailed proof.
Ideally, we require to take the expectation values over many runs according to Eq.~(\ref{QDRIFT-QITE}). In Section \ref{Numerical}, we will see that a simplified version, single-path td-DRIFT-QITE, which samples only a single unitary evolution path instead of a quantum channel, suffices in practice to obtain accurate results when $\Delta t$ is reasonably small.

Randomization brings the optimization to fluctuate, which fluctuation inevitably increases the running time of the circuit. 
To alleviate the fluctuation, we introduce a heuristic alternative to the randomized protocol. In analogy with adaptive VQE~\cite{grimsley2019adaptive,tang2021qubit}, which repetitively chooses the operator with the largest gradient in the previous iteration and adds it to the quantum circuit until it converges or satisfies the termination criteria, we can also deterministically select the term with the largest coefficient $a_i$ in $A$. We note that this deterministic version is a special case of single-path td-DRIFT-QITE, where the random sampler happens to choose the Pauli term with the biggest magnitude at every Trotter step. In our numerical result for the $\mol{LiH}$ molecule in Section~\ref{Numerical}, we observe that compared to its randomized counterpart, this deterministic variant further reduces circuit depth, especially for geometries with strong static correlation. However, due to its deterministic nature, this version may be more sensitive with respect to step size compared to td-DRIFT-QITE. One may need to fine-tune the step size in order to obtain better energy convergence while using shallower circuits.

\section{Measurement reduction and error analysis}

For our methods and the other QITE-based variants described above, one uses the full Hamiltonian $H$ at a single Trotter step to obtain the unitaries. This requires to measure $\bold{S}$ and $\bold{b}$, and then calculate the coefficients of the unitaries by solving $\bold{S}\bold{a} = \bold{b}$ at each time step. 
Intuitively, the number of measurement with the uniform measurement strategy scales asymptotically as $\mathcal{O} (K (L  \nu + \nu^2))$, where $K$ is the number of iteration, $\nu$ is the size of the operator pool for the drifted unitaries, and $L$ is the number of terms of the Hamiltonian. An essential question for the efficient implementation of QITE is how to reduce the sample complexity. In Section~\ref{M_reduction}, we propose a general observable estimation scheme that alleviates the overhead $K$. In Section~\ref{sec:meas_distribute}, we show how to remove the dependence on $L$, and provide a measurement distribution strategy for different observables. In Section~\ref{error_analysis}, we discuss the numerical instability of the matrix inversion, and show its relation to the correlation length of the target problem.

\subsection{Measurement reduction for observable estimation}\label{M_reduction}

Now we discuss the measurement reduction scheme for an arbitrary observable $\hat O$.
We mainly take advantage of the fact that state $\Psi_{k}$ at the time step $k$ is obtained by only adding an operator generated by a single Pauli $P_k$ on $\Psi_{k-1}$.
Hence, we may be able to reuse the estimations of expected values measured from previous rounds to assist the evaluation at the new round. At every new iteration, the state is updated as 
\begin{equation}
\ket{\Psi_k}= e^{-i c_k P_k \Delta t}\ket{\Psi_{k-1}}
\end{equation}
where coefficient $c_k$, $P_k$ correspond to the norm of $a$ and the Pauli operator selected by the drifting algorithm in Section~\ref{algorithm}, respectively. For any observable $\hat{O}$ to be evaluated, the expectation value with respect to a single-step update is
\begin{equation}
\begin{split}
&\bra{\Psi_k}\hat{O}\ket{\Psi_k}=\bra{\Psi_{k-1}}e^{i c_k P_k\Delta t}\hat{O}e^{-i c_k P_k\Delta t}\ket{\Psi_{k-1}}\\& = \bra{\Psi_{k-1}}\hat{O}\ket{\Psi_{k-1}} + i c_k \bra{\Psi_{k-1}}[P_k,\hat{O}]\ket{\Psi_{k-1}}\Delta t +\mathcal{O}(\Delta t^2)
\end{split}\label{MDA}
\end{equation}
Therefore, provided an estimation to $\bra{\Psi_{k-1}}\hat{O}\ket{\Psi_{k-1}}$ in the previous step, measured to a good precision, one may only need to measure $\bra{\Psi_k}[P,\hat{O}]\ket{\Psi_k}$ for the next step. Importantly, as the variance contribution from $\bra{\Psi_k}[P,\hat{O}]\ket{\Psi_k}$ is  suppressed by a small $\Delta t$, the samples needed for the next iteration are not increased greatly. Using this equation recursively, one can trace back to $\ket{\Psi_{0}}$ and throwing away the second-order terms for every iteration, we have 
\begin{equation}
\begin{split}
\bra{\Psi_k}\hat{O}\ket{\Psi_k}\approx \bra{\Psi_{0}}\hat{O}\ket{\Psi_{0}} + 
\sum_{s=1}^k i c_s\bra{\Psi_{s-1}}[P_s,\hat{O}]\ket{\Psi_{s-1}}\Delta t
\end{split}
\end{equation}
In the numerical results in Section ~\ref{Numerical}, we show evidence for the validity of this approximation. Here, we use this first-order approximation of $\bra{\Psi_k}\hat{O}\ket{\Psi_k}$ to obtain the following result for the total number of measurements needed for all time steps.
\begin{prop}
To estimate the observable up to precision $\varepsilon$, the total number of measurements during all time steps scales as   $\mathcal{O} ({ (1+\norm{c}_\infty T)^2}{\varepsilon^{-2}} )$, where $T$ is the total time, $\|c\|_\infty := \max_k c_k$. The optimal measurement distribution strategy is
$n_0=\frac{N_k}{k\norm{c}_\infty\Delta t+1}$ and $n_1=\dots=n_k=\frac{N_k\norm{c}_\infty\Delta t}{k\norm{c}_\infty\Delta t+1}$, where $n_s$ denotes the number of samples for the $s$th iteration and $\Delta t$ is the time step.
\label{prop:meas_reduce}
\end{prop}

In our measurement scheme, the scaling is independent of the number of time steps, therefore it shows an improvement standard the naive measurement scheme in QITE, i.e., measuring observable for each iteration independently, which scales as $\mathcal{O} ({K }{\varepsilon^{-2}})$. We remark that this scheme works for all variants of QITE that add operators to the circuit one after another for different time step. 
More details of the proof can be found in Appendix~\ref{REDUCTION}.

\subsection{Measurement strategy} 
\label{sec:meas_distribute}

In the above section, we have established a general measurement scheme during the imaginary time evolution.
Now, we focus on the number of measurements required for the estimation of $\bold{S}$, $\bold{b}$, and $c$ at each time step.
The coefficients $\bold{a}$ are obtained by $\bold{a} =\tilde{\bold{ S }} ^{-1} \bold{b}$, where $\tilde{\bold{ S }}$ is $\bold{S}$ after truncating small singular values because $\bold{S}$ may be ill-conditioned.
As what we need are the estimation of the coefficients,
it may not be necessary to achieve the same error $\varepsilon$ for every matrix element of $\bold{S}$ and $\bold{b}$. We propose a measurement distribution strategy for different terms in the equation to reduce the variance of $a$.

It is straightforward to check that $\mathrm{Var} (\tilde{\bold{ S }}_{ij}) \leq 1$.
The variance of the unnormalized elements of $\bold{b}$, defined as $\bar{\bold{b}}:= -\Im \bra{\Psi}H\sigma_i\ket{\Psi}$, can be bounded by 
$
\mathrm{Var}(\mathrm{Im}\bra{\Psi}H\sigma_i\ket{\Psi}) \leq \|\bold{h}\|_1^2,
$
where $\|\bold{h}\|_1:= \sum_l h_l$ represents the $l_1$ norm of the absolute sum of Hamiltonian strengths.
Here, use the $l_1$-sampling measurement strategy, and other measurement methods such as classical shadows may be used to reduce the measurement complexity~\cite{huang2020predicting,zhang2021experimental}. 
We can show that the variance of the inverse square root of normalization factor satisfies  
$
    \mathrm{Var}({c}^{-\frac{1}{2}}) \leq \Delta t^2 \|\bold{h}^2\|_1/[c]^3
$,
where $[c]$ represents the expectation of $c$.

Next, we show how to distribute the measurements to control the variance of the solution of the linear equation $\bold{a}$.
Assume that we assign the number of measurements $N_S$, $N_{\bar b}$, and $N_c$ to estimate the matrix $\bold{S}$, vector $\bold{\bar b}$, and the normalization factor $c$ respectively.
To reduce the uncertainty of $\bold{a}=\tilde{\bold{ S }}^{-1}\bold{b}$ to a precision $\varepsilon$, the number of samples to estimate  $\bold{a}$ can be set as 
\begin{equation}
\begin{split}
    N_{S}&\geq 3 \varepsilon^{-2} \nu^2  \norm{\bold{b}}^2_\infty \|\tilde{\bold{ S }}^{-1}\|_F^4  \\
    N_{\bar b}&\geq 3\varepsilon^{-2}  \nu [c]^{-1} \|\bold{h}\|_1^2 \|\tilde{\bold{ S }}^{-1}\|_F^2\\
    N_{c}&\geq 3 \varepsilon^{-2} \Delta t^2[c]^{-3} \|\bold{h}\|_1^2 \|\tilde{\bold{ S }}^{-1}\|_F^2
\end{split}
\end{equation}
where $\|A \|_F:= \sqrt{\tr(A^{\dagger} A)}$ is the Frobenius norm, $\norm{\bold{b}}^2_\infty:=\max_i b_i$, and $\nu$ is the size of the Pauli operator pool.
We can find that the number of samples  is independent of the number of terms $L$ of the Hamiltonian, and thus can be particularly useful for the quantum chemistry problems.
We leave a detailed derivation of the variance and the number of measurements to Appendix \ref{REDUCTION_two}.

\subsection{Correlation length and numerical instability}\label{error_analysis}

Using the sensitivity result for the linear system~(see Appendix \ref{sensitivity_append}), we have the following relationship
\begin{equation}
\delta(\bold{a})=\kappa(\bold{S})(\delta(\bold{S})+\delta(\bold{b})),
\label{sensitivity}  
\end{equation}
where $\delta(\cdot)$ denotes the relative error of the quantity and $\kappa(\cdot)$ denotes the condition number. The condition number of $\bold{S}$ may be large, and thus the small error on the matrix elements will result in a large deviation of $\bold{a}$. Therefore, it is necessary to truncate the small singular values or regularize $\bold{S}$ during evolution. This truncation could lead to the deviation of QITE from the exact imaginary time evolution. We use the notion of correlation length to analyse the error due to 
strong correlation. Molecular systems may exhibit correlations with power-law decay due to long-range correlation between orbitals, and thus we assume that the correlation function satisfies the following inequality~\cite{hastings2006spectral}
\begin{equation}
\langle O_1O_2 \rangle-\langle O_1 \rangle\langle O_2 \rangle\leq \alpha\norm{O_1}\norm{O_2}(1+d(O_1,O_2))^{-1/\xi} \label{inequality_corr_len}
\end{equation}
where $\xi$ is the correlation length,  $d(O_1,O_2)$ is the distance between the supports of observables $O_1$ and $O_2$, and $\alpha$ is a constant. The expectation here is taken with respect to the states along the QITE path. 

In Appendix \ref{sensitivity_append}, we use the inequality in Eq.~(\ref{inequality_corr_len}) to analyze the distribution of the singular values of $\bold{S}$ for the following two case: systems whose correlation length is close to 0 and those whose correlation length is large. Observing the different pattern of the singular value distribution for these two cases, we argue that truncation error is one of the main error sources for systems with large correlation length.

\section{Numerical simulation}\label{Numerical}

In this section, we test the effectiveness of our algorithms on several benchmark molecular Hamiltonians. Specifically, we test td-DRIFT-QITE on $\mol{LiH}$, $\mol{BeH_2}$ molecules, the deterministic version on $\mol{LiH}$, $\mol{N_2}$ molecules and the faithfulness of the approximation in our measurement reduction on $\mol{N_2}$ molecule. At the bond lengths where our calculation reaches chemical accuracy, we demonstrate the effectiveness of depth reduction by counting the number of time steps required. As we will see, these numbers are much smaller than the number of Pauli operators in the UCCSD pool, and hence the depth of our whole calculation is even smaller than that of one time step with original QITE on the same pool. Towards the dissociation limit, our results fail to converge to chemical accuracy. As we show in Fig.~\ref{truncation} of Appendix~\ref{sensitivity_append}, the original QITE with negligible truncation is able to reach chemical accuracy at those bond lengths, which resonate with our error analysis in Section~\ref{error_analysis}. We also show numerical support for the convergence of td-DRIFT-QITE to QITE.

\begin{table}
    \centering

    \begin{tabular}{||c|c|c|c|c|c|c||}
    \hline
    bond lengths/$\AA$ & 0.8 & 1.0 & 1.2 & 1.4 & 1.6 & 1.8\\
    \hline 
    td-DRIFT & 12 & 11 & 9 & 13 & 18 & 27\\ 
    \hline 
    deterministic & 11 & 13 & 14 & 17 & 17 & 18\\
    \hline
    \end{tabular}
    \begin{tabular}{||c|c|c|c|c|c|c||}
    \hline
    bond lengths/$\AA$ & 2.0 & 2.2 & 2.4 & 2.6 & 2.8 & 3.0\\
    \hline 
    td-DRIFT & 31 & 26 & 71 & 87 & 105 & 103\\ 
    \hline 
    deterministic & 21 & 28 & 36 & 46 & 50 & 59\\
    \hline
    \end{tabular}
    \caption{The number of time steps~(circuit layers generated by a single Pauli) needed for single-path td-DRIFT-QITE and its deterministic counterpart to reach chemical accuracy for $\mol{LiH}$ with bond lengths from 0.8 $\AA$ to 3.0 $\AA$. These numbers for circuit depths are comparable to one of the state-of-the-art VQE methods, qubit-ADAPT VQE method~\cite{tang2021qubit}. }\label{LiH}

\end{table}

\begin{figure*}
  \includegraphics[width=\textwidth]{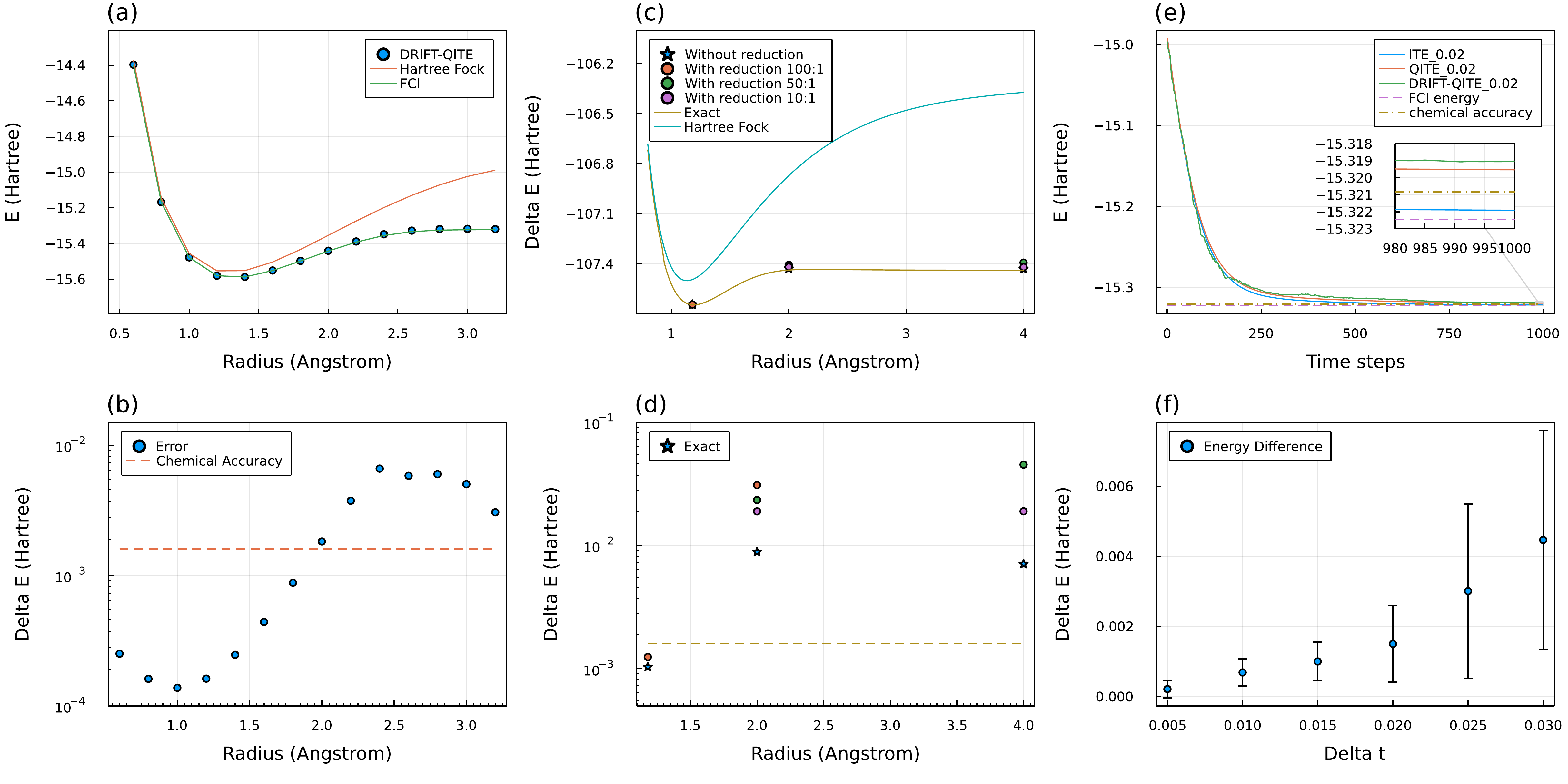}
  \caption{(a) The potential energy surface of $\mol{BeH_2}$ obtained with single-path td-DRIFT-QITE (without measurement reduction). (b) Errors from energies obtained by exact diagonalization~(ED). (c) The potential energy surface of $\mol{BeH_2}$ obtained with the deterministic td-DRIFT-QITE with different ratio of reduction~(e.g., 10:1 means 1 time step without measurement reduction followed by 9 with measurement reduction for every 10 time steps.). (d) Errors from ED energies. (e) Convergence curves of ITE, QITE and single-path td-DRIFT-QITE for $\mol{BeH_2}$ at bond length 3.2$\AA$. (f) Energy discrepancy between td-DRIFT-QITE and QITE and the standard deviation of td-DRIFT-QITE for $\mol{BeH_2}$ at bond length 3.0$\AA$. The Hamiltonian for each configuration is generated using the PySCF package at the basis of STO-3g~\cite{sun2018pyscf} and transformed to Pauli representation using OpenFermion package~\cite{mcclean2020openfermion}, and the ansatz construction and circuit simulations are performed with Yao~\cite{luo2020yao}. }\label{FIGURE}
\end{figure*}

To demonstrate the effectiveness of depth reduction and accuracy of single-path td-DRIFT-QITE, we list the number of time steps needed to reach chemical accuracy in Table~\ref{LiH} for $\mol{LiH}$ molecule with bond lengths ranging from 0.8 $\AA$ to 3.0 $\AA$. The step size is chosen to be $0.16$ for all bond lengths. All of these numbers are much smaller than 176, which is the number of Pauli operator in the corresponding UCCSD pool~(with frozen core implemented). This means our algorithm needs a much shallower circuit for the whole calculation compare to the original QITE circuit with UCCSD ansatz pool in just one time step. In order to confirm the validity of our scheme for larger systems and operator pools, we run td-DRIFT-QITE for $\mol{BeH_2}$ and plot its potential energy surface in Fig.~\ref{FIGURE} (a). The calculation reaches chemical accuracy compared to FCI at STO-3g basis with equal bond lengths of B-H at 0.6, 0.8, 1.0, 1.2, 1.4, 1.6 and 1.8$\AA$. The chemical accuracy is reached for these configurations at time step 31, 57, 53, 52, 37, 73 and 120, respectively, with $\Delta t=0.03$.
Note the step size $\Delta t$ is the same for all the bond length in Fig.~\ref{FIGURE} (a), (b). One can take bigger step size to further reduce depth. Even so these number of time step are much smaller than 456, which is again the number of terms in the UCCSD Pauli operator pool of $\mol{BeH_2}$~(with frozen core implemented). 

The ground state energies obtained for bond lengths at the dissociation limit fails to reach chemical accuracy. We show evidence in Fig.~\ref{FIGURE} (e) that this is not due to the error of drifting approximation. We plot the convergence curves of ITE, QITE, and td-DRIFT-QITE~(average of 10 different sampled paths) for $\mol{BeH_2}$ at bond length 3.2$\AA$. The curve of td-DRIFT-QITE fluctuates around QITE and even converges faster than ITE at some stages. In the end td-DRIFT-QITE converges within 0.5mHa from QITE, 3mHa higher than exact ground state energy. Therefore, drifting approximation accounts for a discrepancy of 0.5mHa, which is well within the need for chemical accuracy, but both fails to reach chemical accuracy because of truncation errors. Here, singular values smaller than the truncation threshold of 0.05 are discarded for both QITE and td-DRIFT-QITE. In Fig.~\ref{truncation} of Appendix~\ref{sensitivity_append}(b), one can see that QITE with negligible truncation is able to reach chemical accuracy. Therefore, if one can prevent the condition number of $\bold{S}$ from exploding by choosing ansatz Pauli operators more carefully, one can in principle reach chemical accuracy using td-DRIFT-QITE with negligible truncation as well.

In order to support our assumption of QITE and proof of bound for Proposition~\ref{prop:bound_prop} in Appendix~\ref{bound_proof}, we show numerical evidence for the bound for convergence of td-DRIFT-QITE to QITE at small $\Delta t$ limit in Fig.~\ref{FIGURE} (f). The convergent energies for td-DRIFT-QITE and QITE are obtained at $\Delta t= 0.005, 0.01, 0.015, 0.02, 0.025$ with the same truncation threshold. The total evolution time for td-DRIFT-QITE at different step size $\Delta t$ are all fixed to be 15. Energies for td-DRIFT-QITE are obtained by running 5 times and taking the average for each bond length. The error bar for each point shows the standard deviation of the 5 energies obtained. As $\Delta t$ decreases, the discrepancies of td-DRIFT-QITE and QITE becomes closer.

As we discussed in Section~\ref{algorithm}, the deterministic variant can avoid the random fluctuation of td-DRIFT-QITE, and therefore will give even shallower circuit in practice. As we can see in Table~\ref{LiH}, this advantage of our deterministic algorithm over the randomized version is more dominant towards the dissociation limit. We test the deterministic algorithm on a system with stronger correlation, $\rm{N_2}$, in Fig.~\ref{FIGURE} (c), (d). At bond length $1.18\AA$, the results reach chemical accuracy within 100 time steps. At bond lengths $2.0\AA$ and $4.0\AA$, the energies converge within 10mHa from the exact result within 400 time steps, while the total number of Pauli operators in the operator pool is 1044. 

In our measurement reduction scheme, an approximation is made in Eq.~(\ref{MDA}). We also test the effect of this approximation on $\rm{N_2}$ for all three bond lengths in Fig.~\ref{FIGURE} (c), (d). A parameter that one can adjust for the measurement reduction protocol is the ratio of the total number of time steps to the number of time steps that do not implement measurement reduction, which we will denote as $\gamma$. For example, $\gamma=10:1$ means 1 time step without measurement reduction followed by 9 with reduction for every 10 time steps of td-DRIFT-QITE. One could probably anticipate that the large $\gamma $ is the worse the approximation will be. Indeed, our result at equilibrium bond length $1.8\AA$ reaches chemical accuracy in 100 time steps even with $\gamma = 100:1$. At bond lengths $2.0\AA$ and $4.0\AA$, the energies converge at about 20mHa away from exact result with $\gamma = 10:1$. These numbers are comparable to the results without measurement reduction and the protocol reduces the measurement overhead by close to a factor of 10 in this case.

\section{Discussion}

In this work, inspired by the idea of random compiling, we proposed a low-depth algorithm for efficient implementation of quantum imaginary time evolution (td-DRIFT-QITE), and a deterministic variant that further reduces the random fluctuation and circuit depth in practice for systems with strong static correlation. Using a drifting protocol on the Pauli terms that generate the QITE ansatz, we remove the circuit depth dependence on the size of the Pauli operator pool. In order to reduce the measurement cost of our algorithms, we utilize the feature of QITE that only a thin circuit layer generated by a single Pauli is added when we measure the corresponding observable at each time step, and this enables to greatly reduce the measurement cost, independent of the number of time steps, showing an advantage in terms of measurement overhead compared to the VQE methods. 
We also remark that there are many great works to reduce the measurement cost for variational algorithms~\cite{kubler2020adaptive,van2021measurement}. 

Our method (td-DRIFT-QITE) serves as a module that can be incorporated into a quantum-classical framework to enhance the capability of the current quantum computers. For example, it can provide a trial state or an initial state in other quantum algorithms, such as in the auxiliary-field quantum Monte Carlo~\cite{huggins2021unbiasing}, ground and excited states preparation~\cite{zeng2021universal,Lin2020Near,lu2021algorithms}, dynamics simulation~\cite{dalmonte2018quantum,sun2021perturbative}, etc. It can also be treated as a quantum solver in the quantum embedding theory, such as density matrix embedding theory~\cite{DMET2012,rubin2016hybrid,li2021toward}, hybrid quantum-classical tensor networks~\cite{yuan2021quantum}, etc.

We could further improve the simulation accuracy when dealing with strong correlation at the dissociation limit. We could consider reducing the truncation error with a more compact basis set than UCCSD operator pool~\cite{zhang2021shallow,fan2021circuit}, which reduces the dimension of the set of linear equations to be solved, and thus the condition number may also be reduced. One can also partition the basis set and traverse all subsets as the number of time steps increases. This effectively constructs multiple reduced operator pools and thus also reducing the measurement cost when constructing the linear equations. We can also search for better schemes for tuning Trotter step size.

\section{Acknowledgement}

The authors thank Yusen Wu, Bujiao Wu, Xiaoming Sun and Xiao Yuan for helpful discussions on relevant topics.

\appendix

\widetext

\section{Error analysis of drifting Hamiltonian and bound for convergence of td-DRIFT-QITE}
\label{bound_proof}


In this section, we first discuss the convergence condition of our algorithm and the upper bound of the simulation error with respect to the number of iteration under certain assumption. Then we consider the scheme of drifting the Hamiltonian itself and argue that it would not work without major modification.

Consider a many-body Hamiltonian $H = \sum_{l=1}^L H_l$. In the original proposal, the authors considered the Trotterized imaginary time evolution, $e^{-H \Delta t} = \prod_l e^{-H_l \Delta t} + \mathcal{O} (\Delta t^2)$, and then approximate the evolution of a pure state $\ket{\Psi}$ under each  $e^{-H_l \Delta t}$  by a  unitary operators as
\begin{equation}
    	\frac{1}{\langle{\Psi|e^{-2H_l \Delta t}|\Psi \rangle}} e^{-H_l \Delta t} \ket{\Psi} \approx e^{-i A \Delta t} \ket{\Psi}. \label{A1}
\end{equation}

The reason why the original proposal considered approximating the Trotterized imaginary time evolution instead of the full time-sliced  $e^{-H \Delta t}$ is to bound the domain size of the unitary operators. 
In particular, if the Hamiltonian is $k$-local, the domain size at the $j$th step is proven to be bounded by $D = \mathcal{O} \left( k C^d \ln ^{d}\left( j L \varepsilon^{-1}\right) \right)
$
qubits, such that the error of the state approximation has a bounded error
$
\|\left|\Psi_{j}\right\rangle-\left|\Phi_{j}\right\rangle \| \leq \varepsilon,
$
for every $\varepsilon > 0$. Here, $d$ is the system dimension, $C$ is an upper bound on the correlation length of the quantum state, {$\ket {\Phi_{ j}}$ } represents the exact state and $\ket {\Psi_{j}}$ represents the approximated state after the nonunitary operations at the $j$th step.

A major caveat of QITE is that the domain size $D$ that the unitaries act on might increase drastically under evolution. Even for the local Hamiltonian and product initial states, the computational cost scales polynomial on the number of terms of the Hamiltonian and exponential to the correlation length. For the quantum chemistry problems, the domain size is still unknown. Instead, the unitary operator was suggested to choose from the UCCSD operator pool in the step-merged QITE.

Now we discuss the error of our td-DRIFT-QITE scheme and the condition under which Proposition~1 holds.
Let $A^{(j)}$ denote the $A$ operator obtained from the original QITE at the $j$th Trotter step and $U_j=e^{iA^{(j)}\Delta t}$. 
Given a decomposition $A^{(j)} =\sum_{i} a_i\sigma_i$, we define a random variable $X^{(j)} = \norm{\bold{a}}_1\sigma_i $ with probability $p_{j,i} = \frac{a_i}{\norm{\bold{a}}_1}$ and $V_j := U_{j,i} =e^{iX^{(j)}\Delta t}$. 
Define $\Xi_{j}=U_j...U_1$ and $\Lambda_j = \E(V_j|\Lambda_{j-1})\Lambda_{j-1}$ with $\Lambda_1=E(V_1)$, our goal is to show that the conditions the QITE evolution has to satisfy in order to obtain
 \begin{equation}
 \norm{\Lambda_K-\Xi_K}\leq \mathcal{O}(\frac{1}{K})  
 \end{equation}
where $\norm{.}$ is the operator norm.
 
We first establish the approximation error at the single step. We further denote $\tilde{U}_j$ to be the counterpart of $U_j$ but given the previous operations being $\Lambda_{j-1}$. We note that 
 \begin{equation}
 \begin{split}
   \|\tilde{U}_j  - \E(V_j|\Lambda_{j-1}) \| &=\| \mathbb{E}[\exp (-\mathrm{i} X^{(j)}\Delta t)]-\mathbb{I}+\mathrm{i} X^{(j)}\Delta t]+(\mathbb{I}-\mathrm{i} \mathbb{E}[X^{(j)}\Delta t]-\exp (-\mathrm{i} \mathbb{E}[X^{(j)}]\Delta t)) \| \\
& \leq \mathbb{E}\|\exp (-\mathrm{i} X^{(j)}\Delta t)-\mathbb{I}+\mathrm{i} X^{(j)}\Delta t\|+\|\exp (-\mathrm{i} \mathbb{E}[X^{(j)}\Delta t])-\mathbb{I}+\mathrm{i} \mathbb{E}[X^{(j)}\Delta t]\| \\
& \leq \frac{1}{2} \mathbb{E}\|X^{(j)}\|^{2}\Delta t^2+\frac{1}{2}\|\mathbb{E}[X^{(j)}]\|^{2}\Delta t^2 \leq \mathbb{E}\|X^{(j)}\|^{2}\Delta t^2=( \|a^{(j)}\| \Delta t)^{2}
 \end{split}\label{single_UV_bound}
 \end{equation}
Here we have used the results derived in Ref.~\cite{chen2021concentration}.
 
We will now make the premise that QITE is faithful enough such that it converges to states near the ground state wherever it starts. Therefore the unitaries generated by QITE can be viewed as a vector field whose divergence is negative in the vicinity of the ground state and 0 everywhere else. So even though td-DRIFT-QITE deviates slightly from QITE in one step, but it will still follow the convergence path of QITE in the next step starting from that new point. Now with a QITE vector field converging towards a center, we make the following assumption:
\begin{equation}
\frac{\sum_{j=1}^K\norm{\tilde{U}_j-U_j}}{K}\sim \norm{U_1-\mathbb{E}(V_1)}.  \label{ASSUMPTION}
\end{equation}
where $U_1=\tilde{U}_1$ the equivalence is with respect to scaling in function of $K$. Here That is to say, the vector field from QITE tends to concentrate as it gets closer to the ground state, i.e., on average, $\tilde{U}_j$s' deviation from the original QITE's convergence direction $U_j$ does not accumulate.

Now we can bound the relative error with respect to the ideal unitary evolution as 
 \begin{equation}
 \begin{split}
 \norm{\Lambda_K-\Xi_K}  &\leq \|U_K \Xi_{K-1} - \tilde{U}_K \Lambda_{K-1}  \|  + \|\tilde{U}_K \Lambda_{K-1} - \E(V_K|\Lambda_{K-1}) \Lambda_{K-1}\|  \\
 &\leq \| U_K \Xi_{K-1} - U_K\Lambda_{K-1} \| + \| U_K\Lambda_{K-1} - \tilde{U}_K \Lambda_{K-1} \|  + \|\tilde{U}_K \Lambda_{K-1} - \E(V_K|\Lambda_{K-1}) \Lambda_{K-1}\| \\ & \leq \| \Xi_{K-1} - \Lambda_{K-1} \| + \| U_K - \tilde{U}_K \|+\|\tilde{U}_K - \E(V_K|\Lambda_{K-1})\| \\
 &\leq \sum_{j}^K   (\|U_j - \tilde{U}_{j}\| + \| \tilde{U}_j - \E(V_j|\Lambda_{j-1}))\| )\\
 &\sim \sum_{j}^K   \|\tilde{U}_j  - \E(V_j|\Lambda_{j-1}) \| 
 \end{split}\label{Bounding_the_whole}
 \end{equation}

 where we have used the distance invariance under the unitary transformation and triangular inequality, and the last line comes from about assumption in Eq.~(\ref{ASSUMPTION}). Because they are of the same order, we only need to bound one of them as we did in Eq.~(\ref{single_UV_bound}). Plugging this bound into Eq.~(\ref{Bounding_the_whole}), we have the convergence in terms of step size bounded as 
\begin{equation}
    \norm{\Lambda_K-\Xi_K} \leq  T \Delta t \max_j  \| a^{(j)} \|^{2} = \mathcal{O} (\frac{1}{K})
\end{equation}

Therefore, we complete the proof of Proposition \ref{prop:bound_prop}.

An alternative approach that the reader may think of is to drift the Trotterized imaginary time operator itself.
However, it is not obvious how to define a mixing quantum operation as that in the original qDRIFT scheme without ancilla qubits.
On the other hand, drifting the unitary approximation method is state-dependent and it might introduce a large simulation error due to the exponentially vanishing normalization factor $c_j := \tr[e^{- 2H \Delta t }\rho ]$. 

If one considers implementing the nonunitary imaginary time operator by the joint evolution $\exp(-i \Delta t H \otimes Y_a)$ and post-processing the measurement outcome on the ancilla~\cite{kosugi2022imaginary}, the imaginary time operation is achieved by post-processing the measurement outcome to be zero as 
\begin{equation}
M_0 = \langle{0_a|\exp(-i \Delta t H \otimes Y_a) | 0_a} \rangle \propto \exp(- \Delta t H  ). 
\end{equation} 
We can find that the imaginary time operator is realized by the real time operator, and thus the mixing channel of the real time evolution is similar to the discussion in our work.

\section{Measurement cost after the reduction scheme}\label{REDUCTION}

Our measurement reduction scheme is trying to reuse estimations of expected values from previous rounds to assist the evaluation at the new round. In this section, we discuss the number of measurements needed when we use our measurement reduction scheme.
The main results are presented in Proposition \ref{prop:meas_reduce}. We show the proof here.

We first denote that the variance of a single sample taken for $ \bra{\Psi_{0}}\hat{O}\ket{\Psi_{0}} $ and any $\bra{\Psi_{s-1}}[P_s,\hat{O}]\ket{\Psi_{s-1}}$ is $\sigma_0^2$ and $\sigma_s^2$ respectively. We also denote the total number of the samples we need by $N_k=\sum_{i=0}^k n_i$, where $n_0$ is the number of sample for $ \bra{\Psi_{0}}\hat{O}\ket{\Psi_{0}} $ and $n_s$ is for $\bra{\Psi_{s-1}}[P_s,\hat{O}]\ket{\Psi_{s-1}}$. {
In the main text, we have shown that the expectation is approximated by its first-order expansion as $\bra{\Psi_k}\hat{O}\ket{\Psi_k}\approx \bra{\Psi_{0}}\hat{O}\ket{\Psi_{0}} + \sum_{s=1}^k i c_s\bra{\Psi_{s-1}}[P_s,\hat{O}]\ket{\Psi_{s-1}}\Delta t$
}

Then we have the variance
\begin{equation}
\begin{split}
&\mathrm{Var}(\bra{\Psi_k}\hat{O}\ket{\Psi_k})
\\
&\approx\mathrm{Var}(\bra{\Psi_{0}}\hat{O}\ket{\Psi_{0}}) + 
\sum_{s=1}^k c_s^2 \mathrm{Var}(\bra{\Psi_{s-1}}[P_s,\hat{O}]\ket{\Psi_{s-1}})\Delta t^2\\
&= \sigma_0^2/n_0 + \sum_{s=1}^k  c_s^2 \sigma_s^2 \Delta t^2/n_s \\
&\leq 
\sigma_0^2/n_0 + \norm{c}_\infty^2 \sum_{s=1}^k   \sigma_s^2 \Delta t^2/n_s
\end{split}
\end{equation}
One can set $\sigma_0^2/n_0 + \norm{c}_\infty^2 \sum_{s=1}^k   \sigma_s^2 \Delta t^2/n_s \leq \varepsilon^2$ to make $\mathrm{Var}(\bra{\Psi_k}\hat{O}\ket{\Psi_k})
\leq \varepsilon^2 $. By Cauchy's inequality, we have
\begin{equation}
\begin{split}
    \varepsilon^2 N_k &\geq(\sigma_0^2/n_0 + \norm{c}_\infty^2 \sum_{s=1}^k   \sigma_s^2 \Delta t^2/n_s)(n_0+n_1 \dots +n_k)\\
    &\geq (\sigma_0+\norm{c}_\infty \sum_{s=1}^k\sigma_s \Delta t)^2=\sigma^2(1+\norm{c}_\infty k \Delta t)^2
\end{split}
\end{equation} 
where $\sigma={\max}_s  \sigma_{s}$. Hence the total number of measurements needed is
\begin{equation}
    N_k\geq \frac{\sigma^2(1+\norm{c}_\infty k \Delta t)^2}{\varepsilon^2} =\frac{\sigma^2(1+\norm{c}_\infty t)^2}{\varepsilon^2}
\end{equation}
The equality holds if and only if $n_0=\frac{N_k}{k\norm{c}_\infty\Delta t+1}$ and $n_1=\dots=n_k=\frac{N_k\norm{c}_\infty\Delta t}{k\norm{c}_\infty\Delta t+1}$.
We thus complete the proof.

\section{Measurement strategies for $\bold{S}$, $\bold{b}$ and $c$} \label{REDUCTION_two}

In this section, we discuss the number of measurements required for the estimation of $\bold{S}$, $\bold{b}$ and $c$. We show how to distribute the measurements  to control the variance of $a$, the solution of the linear equation $\bold{S} \bold{a} = \bold{b}$.

Up to the first-order expansion, we have $\bold{b}=-\Im \bra{\Psi}H\sigma_i\ket{\Psi}/\sqrt{c} $ , where $c\approx \bra{\Psi}(1-2H\Delta t)\ket{\Psi}$.
Then the coefficients of the unitaries that approximates the imaginary time evolution is determined by
\begin{equation}
     \bold{S} \bold{a}=\bold{b}.
\end{equation}

Note that each matrix element $\bold{S}_{ij}=\bra{\Psi}P_iP_j\ket{\Psi}$ where $\{P_i\}_i$ are all Pauli operators. The variance of elements of $\bold{S}$ 
\begin{equation}\label{f_s}
        \mathrm{Var}(\bold{S}_{ij})
        \leq 1
\end{equation}
Here we use the $l_1$-sampling method to measure the unnormalized elements of $\bold{b}$, defined as $\bar{\bold{b}}:= -2\Im \bra{\Psi}H\sigma_i\ket{\Psi}$. We can show that the variance is bounded by 
\begin{equation}
\begin{split}
\mathrm{Var}(\mathrm{Im}\bra{\Psi}H\sigma_i\ket{\Psi}) \leq \|\bold{h}\|_1^2
\end{split}
\end{equation}
where $\|\bold{h}\|_1:= \sum_l h_l$ represents the $l_1$ norm of the absolute sum of Hamiltonian strengths.
Other measurement methods may be used to improve the variance, and we refer to Ref.~\cite{huang2020predicting,wu2021overlapped,zhang2021experimental} for more details.

Similarly, we have $\mathrm{Var}(c) \leq 4 \Delta t^2 \|\bold{h}\|_1^2$.
Note the fact that $\mathrm{Var}({c}^{-\frac{1}{2}})\approx \mathrm{Var}(c)/{4[c]^3}$, which can be derived from the Taylor approximation as
\begin{equation}
    \begin{split}
        \mathrm{Var}({c}^{-\frac{1}{2}}) &\approx\mathrm{Var}({[c]^{-\frac{1}{2}}}+\frac{1}{2\sqrt{[c]^3}}(c-[c]))  = \frac{1}{4[c]^3}\mathrm{Var}(c).
    \end{split}
\end{equation}
The variance of the inverse square root of normalization factor thus satisfies  
\begin{equation}\label{f_c}
    \mathrm{Var}({c}^{-\frac{1}{2}}) \leq \Delta t^2 \|\bold{h}^2\|_1/[c]^3
\end{equation}
where we denote $[c]$ as the expectation of $c$.

The variance of $\bold{b}_i$ may be approximated by
\begin{equation}\begin{aligned}
    \mathrm{Var}(\bold{b}_i) \approx [{c}]^{-1} \mathrm{Var}(\bold{\bar b}_i) + [\bold{\bar b}_i]^2\mathrm{Var}({c}^{-\frac{1}{2}}) \\
    = [{c}]^{-1} \mathrm{Var}(\bold{\bar b}_i) + \frac{[\bold{\bar b}_i]^2}{4[c]^3} \mathrm{Var}(c)\\
    \end{aligned}
\end{equation}
Here, we make the assumption that $\bold{\bar b}$ and $c$ are nearly uncorrelated.
From the above derivation,  the variance of the element from a single measurement can be bounded as
\begin{equation}\begin{aligned}
    \mathrm{Var}(\bold{b}_i) 
    \leq (1 + \Delta t^2 [c]^{-2} ) [c]^{-1} \|\bold{h}\|_1^2
    \end{aligned}
\end{equation}

Define the regularized inverse $\tilde{\bold{ S }}^{-1}:=[ (\bold{S} +\lambda \bold{Id})]^{-1}$ of coefficient matrix $\bold{S}$ with the regularization parameter $\lambda \geq 0$, 
Define the variance of $\bold{a}=\tilde{\bold{ S }}^{-1}\bold{b}$ 
as $\sigma^2=\sum_k \mathrm{Var}(a_k)$. 
Again, we assume that  the measurements of elements in $\bold{S}$ and $\bold{b}$ are independently and small. The variance of $\bold{a}$ can be expressed as 
\begin{equation}
    \sigma^2=\sum_{k,l}\alpha_{kl} [\mathrm{Var}(\tilde{\bold{ S }})]_{kl}+\sum_{k}\beta_k [\mathrm{Var}(\bold{b})]_k
\end{equation}
where
\begin{equation}
\begin{split}
    &\alpha_{kl}:=\sum_{i,j}
    [\tilde{\bold{ S }}^{-1}]_{i k}^2
    [\tilde{\bold{ S }}^{-1}]_{l j}^2
    [\bold{b}]_l^2,
    \\
    &\beta_k:=\sum_l[\tilde{\bold{ S }}^{-1}]_{k l}^2.
\end{split}
\end{equation}
See Refs.~\cite{van2021measurement,lefebvre2000propagation} for details.

We have $\sigma^2=\sigma_M^2+\sigma_b^2$, where
\begin{equation}
    \sigma_M^2:=\sum_{k,l}\alpha_{kl} \mathrm{Var}(\tilde{\bold{ S }}_{kl}), 
    \sigma_b^2:=\sum_{k}\beta_k \mathrm{Var}(\bold{b}_k)
\end{equation}

Now, we discuss the estimation of the solution of the linear systems.
Assume that we assign the number of measurements $N_S$, $N_{\bar b}$, and $N_c$ to estimate the matrix $\bold{S}$ , vector $\bold{\bar b}$, and the normalization factor $c$ respectively.
We have the following relation 
\begin{equation}
\begin{split}
    \sigma^2_M\leq 
    & \frac{\nu^2}{N_S} \norm{\bold{b}}_\infty^2 \sum_{i,k}
    [\tilde{\bold{ S }}^{-1}]_{i k}^2
    \sum_{l,j}
    [\tilde{\bold{ S }}^{-1}]_{l j}^2\\
    =&  \frac{\nu^2}{N_S} \norm{\bold{b}}^2_\infty\norm{\tilde{\bold{ S }}^{-1}}_F^4.\\
    \label{eq:sigma_M}
\end{split}
\end{equation}
The first line of Eq.~\eqref{eq:sigma_M}  used the fact that the variance of the element of $M$ is
$
    \mathrm{Var}(\tilde{\bold{ S }}_{ij})=\mathrm{Var}(\bold{S}_{ij} )\leq 1.
$, and the vector infinity norm is defined $\norm{\bold{b}}^2_\infty:= \max_i b_i$.
The second line of Eq.~\eqref{eq:sigma_M} is simplified using the Frobenius norm $\|A\|_F = \sqrt{ \tr[A^{\dagger} A]}$. 

Similarly, we have  
\begin{equation}
\begin{split}
    \sigma^2_b\leq   
    &   (\frac{\nu}{N_{\bar b}}+\frac{\Delta t^2 [c]^{-2}  }{N_c}) [c]^{-1} \|\bold{h}\|_1^2  
    \sum_{k,l}[\tilde{\bold{ S }}^{-1}]_{k l}^2 \\
    & \leq 
      (\frac{\nu}{N_{\bar b}} +\frac{\Delta t^2 [c]^{-2}}{N_c}   ) [c]^{-1} \|\bold{h}\|_1^2 \norm{\tilde{\bold{ S }}^{-1}}_F^2. 
\end{split}
\end{equation}

The variance of $\bold{a}$ under this measurement setting satisfies
\begin{equation}
\begin{split}
    \sigma^2\leq& 
     \norm{\tilde{\bold{ S }}^{-1}}_F^2 \left( \frac{\nu^2 \norm{\bold{b}}^2_\infty \norm{\tilde{\bold{ S }}^{-1}}_F^2}{N_S} + \frac{\nu   [c]^{-1} \|\bold{h}\|_1^2}{N_{\bar b}} + \frac{ \Delta t^2[c]^{-3} \|\bold{h}\|_1^2}{N_c} 
       \right)
       \label{eq:var_tot}
\end{split}
\end{equation}

To reduce the uncertainty of the estimation $\bold{a}=\bold{\tilde{S}}^{-1}\bold{b}$ to a target precision $\varepsilon$, we can set each component in Eq.~\eqref{eq:var_tot} to be less than $\varepsilon^2/3$.
The number of samples to estimate the  $\bold{a}$ is upper bounded by
\begin{equation}
\begin{split}
    N_{S}&\geq 3 \varepsilon^{-2} \nu^2  \norm{\bold{b}}^2_\infty \norm{\tilde{\bold{ S }}^{-1}}_F^4  \\
    N_{\bar b}&\geq 3\varepsilon^{-2}  \nu [c]^{-1} \|\bold{h}\|_1^2 \norm{\tilde{\bold{ S }}^{-1}}_F^2\\
    N_{c}&\geq 3 \varepsilon^{-2} \Delta t^2[c]^{-3} \|\bold{h}\|_1^2 \norm{\tilde{\bold{ S }}^{-1}}_F^2
\end{split}
\end{equation}
which suffices to reach the precision $\varepsilon$.

\section{Derivation of Eq.~(\ref{sensitivity}) and effects of truncation} \label{sensitivity_append}

Assume both sides of Eq.~\ref{equation} are perturbed due to truncation errors or insufficient number of measurements, one has the following expression:
\begin{equation}
(\bold{S}+\varepsilon\bold{S_e})\bold{a}(\varepsilon) = \bold{b}+\varepsilon\bold{b_e}
\end{equation}
Now we take a derivative on both sides with respect to $\varepsilon$:
\begin{equation}
(\bold{S}+\varepsilon\bold{S_e})\dot{\bold{a}}(\varepsilon)+\bold{S_e}\bold{a}(\varepsilon)=\bold{b_e}
\end{equation}
We evaluate this at $\varepsilon = 0$, thus obtain
\begin{equation}
(\bold{S})\dot{\bold{a}}(0)+\bold{S_e}\bold{a}(0)=\bold{b_e}.    
\end{equation}
Taking the inverse of $\bold{S}$:
\begin{equation}
\dot{\bold{a}}(0)=(\bold{S})^{-1}(\bold{b_e}-\bold{S_e}\bold{a}(0)).
\end{equation}
Now we perform a Taylor expansion at $\varepsilon=0$:
\begin{equation}
\begin{split}
\bold{a}(\varepsilon)&=\bold{a}(0)+\varepsilon\dot{\bold{a}}(0)+\mathcal{O}(\varepsilon^2)\\
&=\bold{a}(0)+\varepsilon(\bold{S})^{-1}(\bold{b_e}-\bold{S_e}\bold{a}(0))+\mathcal{O}(\varepsilon^2)
\end{split}
\end{equation}
Taking the norm on both side, we obtain the relative error of $\bold{a}$:
\begin{equation}
\begin{split}
&\frac{\norm{\bold{a}(\varepsilon)-\bold{a}(0)}}{\norm{\bold{a}(0)}}=\varepsilon\frac{\norm{(\bold{S})^{-1}(\bold{b_e}-\bold{S_e}\bold{a}(0))}}{\norm{\bold{a}(0)}}+\mathcal{O}(\varepsilon^2)\\
&\leq \norm{(\bold{S})^{-1}}\norm{\bold{S}}(\frac{\norm{\varepsilon \bold{b}_e}}{\norm{\bold{b}}}+\frac{\varepsilon\norm{\bold{S}_e}}{\norm{\bold{S}}})+\mathcal{O}(\varepsilon^2),
\end{split}
\end{equation}
where $\norm{(\bold{S})^{-1}}\norm{\bold{S}}$ is just the condition number of $\bold{S}$ and $\frac{\norm{\varepsilon \bold{b}_e}}{\norm{\bold{b}}}$, $\frac{\varepsilon\norm{\bold{S}_e}}{\norm{\bold{S}}}$ are the relative errors of $\bold{b}$, $\bold{S}$ respectively.

\begin{figure}[ht]
    \centering
    \includegraphics[height=4.2 cm]{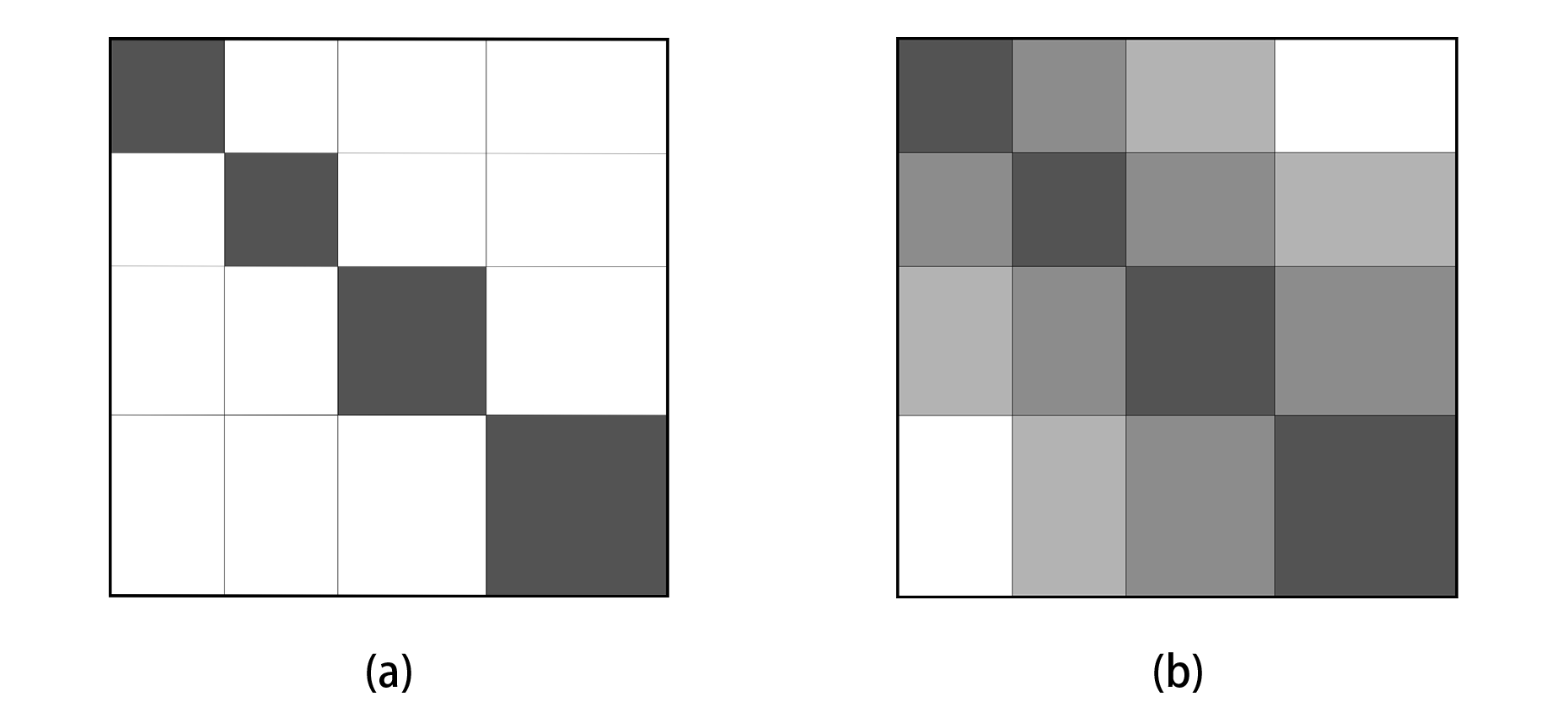}
    \caption{Approximate magnitude of matrix elements of $\bold{S'}$ for different correlation length. Darker colors indicate the values in the corresponding entries have larger magnitudes.}
    \label{corr_len}
\end{figure}

Here we present our arguments and numerical results for the error analysis we mentioned in Section IV. In analogy with Eq.~(\ref{inequality_corr_len}), we define the correlation matrix $\bold{S'}$ and assume it exhibits power-law decay correlation as
\begin{equation}
\bold{S'}_{ij}:=\text{Re}(\langle P_i P_j \rangle- \langle P_i \rangle \langle P_j \rangle )\leq {\alpha}{(1+d(P_i,P_j))^{-1/\xi}}. \label{correlation_length}
\end{equation}
When the correlation length $\xi$ is sufficiently small, the inequality above implies that nonzero values of $d(P_i,P_j)$s will result in $\bold{S}'_{ij}$s that are close to 0. On the other hand, $\bold{S}'_{ij}$ will be upper bounded by $\alpha$ if the corresponding $d(P_i,P_j)=0$. Therefore, one can order the indices of $\bold{S}'$ such that its block-diagonal entries are all upper-bounded by $\alpha$ and others approximately 0, as shown in Fig.~\ref{corr_len}(a). 
This matrix is well-approximately by a matrix with rank equal to the number of blocks $n_b$. The nonzero singular values of this matrix are approximately $\alpha$ times the sizes of the diagonal blocks. Note that the term $\langle P_i \rangle \langle P_j \rangle$ has rank $1$. Hence, $\bold{S}=[\text{Re}\langle P_i P_j \rangle]$ is well approximated by a matrix of rank at most $n_b+1$.  
The distribution of singular values of $\bold{S}$ will be similar to those of $\bold{S'}$, concentrating around a few values. Small deviations from the upper bound $\alpha$ in the block-diagonal terms will give few nonzero singular values other than the $n_b+1$ ones. Therefore, a small truncation will have a negligible effect on the solution of the linear system of equations. As the correlation length $\xi$ increases, $\bold{S'}$ becomes more complicated. It can be well-approximated by a block matrix whose entries' magnitude increases from nondiagonal blocks to diagonal blocks, taking several values instead of merely $0$ and $\alpha$, as shown in Fig.~\ref{corr_len}(b). The distribution of the singular values of this matrix will be more smooth and less concentrated. In this case, the deviation from the exact equality in Eq.~(\ref{correlation_length}) will give larger corrections to the singular values. Therefore, truncation to the singular values will result in larger errors for systems with stronger static correlation. 

\begin{figure}[t]
    \centering
    \includegraphics[height=9.5cm]{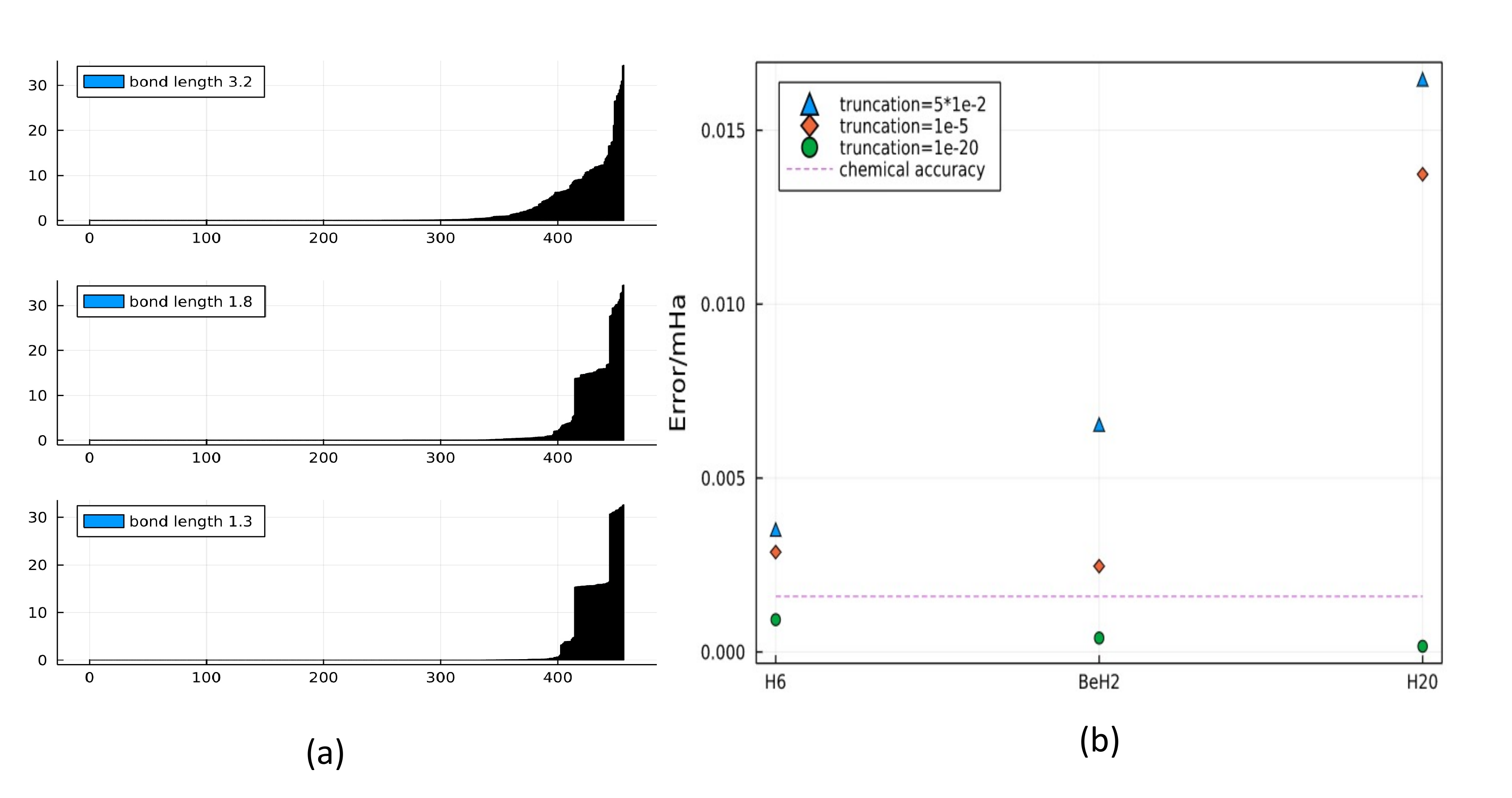}
    \caption{(a). The magnitudes of the singular values of $\bold{S}$ given different bond length for $\mol{BeH_2}$. The closer to equilibrium position~(bond-length 1.3) the molecule is, the smaller the correlation length is, the more magnitudes of the singular values concentrates on a few values as we analyzed in the main text. There are 456 singular values in total, the same as the size of the Pauli operator pool for the UCCSD ansatz of $\mol{BeH_2}$, and they are arranged in ascending order in their magnitudes. $\bold{S}$ is taken at step 200 with $\Delta t=0.03$ and energies nearly converged.
    (b). QITE with different truncation of singular values. The QITE energies are obtained for 3 different molecules, $\mol{BeH_2}$ at bond length 2.4, $\mol{H_6}$ and bond length 2.4 and $\mol{H_2O}$ at bond length 2.0. Energies with truncation of $10^{-20}$ reaches chemical accuracy in all 3 cases. Note that this figure is not meant to compare the accuracy across different molecules as the magnitude of singular values are different for every molecule, while we use the same truncation for them.}
    \label{truncation}
\end{figure}

As we see in Fig.~\ref{truncation}(a), the "tail" of the distribution of the singular values of $\bold{S}$ elongates as correlation length of the system increases. Therefore, truncation these "tails" will inevitably bring larger errors for systems with stronger static correlation. $\bold{S}$ of td-DRIFT-QITE and its deterministic variant usually have very large condition numbers because it is generated by much fewer Pauli ansatz operators, therefore the necessity to truncate the singular values in our algorithms causes errors with the convergence energy. QITE can endure smaller truncation than td-DRIFT-QITE. In Fig.~\ref{truncation}(b) we show the effect of truncation with some examples. QITE with negligible truncation converges to chemical accuracy even towards the dissociation limit. This is to argue that controlling the condition number is essential in order to improve the accuracy of our low-depth algorithms.

\section{Comparison of VQE and QITE variants}
\label{measure_compare}

As previously mentioned in the introduction, variation algorithms, VQE in particularly, have been widely explored to study the ground state of a quantum system in NISQ era. Now we make some comparison between QITE variants~(mainly our algorithm and smQITE) and VQE methods. 

VQE is a hybrid quantum-classical  algorithm in which the quantum device is used to prepare a circuit ansatz and the classical computer performs optimization to find the parameters for the quantum ansatz such that the parameterized state is close to the unknown ground state. There is dramatic progress in experimental and theoretical study of VQE recently~\cite{peruzzo2014variational,shen2017quantum,o2016scalable,hempel2018iontrap,nam2020ground,XU2021,endo2020prl, google2020hartree,grimsley2019trotterized,grimsley2019adaptive,fan2021circuit,QCC,qcc_emb,zhang2020lowdepth,liu2021variational}. Meanwhile, other efficient implementation of variational algorithms for real and imaginary-time evolution has been extensively explored~\cite{gomes2021adaptive, yao2021adaptive, benedetti2021hardware, barison2021efficient, lin2021real}. However, several problems exist which may limit the power of VQE. First, the high-dimensional optimization of parametric quantum circuits is proved to be \rm{NP}-hard in the worst-case scenario~\cite{bittel2021training}. Second, the way to estimate the expressive power of the ansatz to approximate the unknown ground state is not clear~\cite{haferkamp2021linear, holmes2022connecting}. Therefore, it could be challenging to find an ansatz that is shallow enough and free of barren plateau problem. Third, the measurement overhead needed to obtain the gradient of the parameters is linear in terms of iteration number and the size of the operator pool. 

QITE naturally avoids the optimization problem by construction.
For the second problem, UCCSD operator pool is  introduced for VQE, which is  one of the most popular VQE ansatz. Just like QITE as discussed in our work and \cite{gomes2020efficient}, the circuit depth for VQE is also linear in terms of the size of the operator pool. However, the circuit depth of the original QITE also scales linearly the number of Trotter steps. In order to tackle this problem, \cite{gomes2020efficient} proposes to merge terms across different steps that are generated by the same Pauli operators. This relaxes the depth requirement of QITE by eliminating the circuit depth dependency on the number of Trotter steps, which makes it analogous to UCCSD-VQE in terms of circuit construction. However, the ordering of the UCCSD operators in the ansatz remains a problem~\cite{grimsley2019trotterized} as well as the error due to shuffling noncommuting Pauli terms during the merge. In contrast to the way smQITE reduces the depth, our algorithm, td-DRIFT-QITE and its deterministic version, removes the depth dependence on the size of the Pauli operator pool. By construction, our algorithms naturally present an ordering for the ansatz operators and does not suffer from errors due to shuffling noncommuting terms. This is also the feature of ADAPT-VQE and its variants. Meanwhile, as shown in our numerical result, our algorithm often requires a circuit that is much shallower than UCCSD-VQE to obtain satisfactory results and reach chemical accuracy for geometries with mostly dynamic correlation. 

For the third problem, first-order derivatives are usually needed from the quantum computer in order to perform a faster optimization in the VQE settings. However, the overhead for measuring first-order derivatives is higher than measuring zero-order information. Using the parameter-shift rule~\cite{li2017hybrid,mitarai2018quantum,schuld2019evaluating}, the algorithm measures two terms with different parameter shifts in order to obtain the derivative with respect to a single parameter. This overhead is the same as our procedure of measuring $\bold{S}$ and $\bold{b}$ in QITE. The gradient is needed for every VQE iteration, hence, the measurement cost for a single iteration of VQE is close to that of a single step of QITE. With the measurement reduction scheme we proposed, one will be able to edge out the VQE measurement overhead given equal rounds of iterations. It is worth mentioning that due to truncation errors, the precision of our low-depth algorithm has not been able to match that of ADAPT-VQE variants for benchmark molecules at the dissociation limit. However, ADAPT-VQE methods usually requires many more total iterations to converge than even UCCSD-VQE. With our measurement reduction scheme, one could achieve advantage in terms of total time cost when running the algorithm and hence may be preferred when combining with other classical quantum chemistry methods.

\bibliographystyle{unsrt}
\bibliography{main}

\end{document}